\documentstyle[11pt]{article}
\begin{document}
\pagestyle{empty}
\begin{center}
{\Large Mesonic Contribution to the Compton Scattering Amplitude
for Heavy Nuclei\footnote{Work supported by Deutsche Forschungsgemeinschaft
(contract Schu 222/13) and \\
\hspace*{0.5cm} Deutscher Akademischer Austauschdienst
(reference D/94/06221)}
 } \\
\vspace{0.5cm}
M.-Th.\,H\"utt\footnote{e-mail: HUETT@UP200.DNET.GWDG.DE}
and A.I.\,Milstein\footnote{permanent address: Budker Institute of
Nuclear Physics, 630090 Novosibirsk, Russia, \\
\hspace*{0.5cm} e-mail: MILSTEIN@INP.NSK.SU}  \\
 \footnotesize
       II. Physikalisches Institut der Universit\"at G\"ottingen,
       G\"ottingen, Germany
\end{center}
\date{}

\parindent0em
\parskip1.5ex plus 0.5ex minus 0.5ex

\sloppy
\renewcommand{\textfraction}{0.05}
\renewcommand{\topfraction}{0.99}
\renewcommand{\bottomfraction}{0.99}
\newtheorem{abb}{Figure}
\newcommand{\bi}[1]{\bibitem{#1}}

\begin{abstract}
The contribution of mesonic exchange currents to nuclear Compton
scattering is investigated within the framework of a Fermi gas model
of nuclear matter in the non-relativistic limit. The
additional interaction between the nucleons is
accounted for by including two- and three-body diagrams.
As a test of this model, the enhancement constant $\kappa $ is calculated.
The full correlators for the central and tensor part of the
nucleon-nucleon interaction due to pion exchange are obtained and the
energy dependence of the amplitude is investigated. The contribution of the
$\Delta $-excitation to the mesonic part of the Compton amplitude
is calculated explicitely
using an effective Hamiltonian in the static limit.
\end{abstract}



\newpage
\pagestyle{plain}
\pagenumbering{arabic}
\section{Introduction}
    At present nuclear Compton scattering at intermediate energies
    is a highly advanced method to study nuclear and subnuclear degrees
    of freedom. Important information can be obtained on the structure of
    the free nucleon and possible modifications of its properties due to
    binding.
    Compton scattering has
	been the subject of several recent theoretical and
	experimental articles
	\cite{helium-lund,carbon-lund,kappa,brown1,weise1}.
	One of the most promising subjects, which can be
    investigated in this way, is the contribution of mesonic exchange currents
    to the Compton amplitude.
    This contribution is most important
	in the region between the energies of collective nuclear states
	and the $\Delta $-excitation of the nucleon, overlapping with both at
	low and high energies, respectively. Experimentally it is a
	tedious task to separate these contributions,
	which entails that already for the analysis of the data important
	theoretical techniques are applied, such as sum rules and
	dispersion relations
	(see e.g. \cite{baumann,kappa,Hayward}).
	On the other hand, the theoretical description
	involves numerous different methods,
	depending on the energy region and the choice of nucleus (see
	\cite{Riska} and references therein).

	The experimental situation has improved significantly in the last
	years. The accuracy of Compton scattering experiments is now good enough
	to reliably extract some contributions to the scattering amplitude
	\cite{schelhaas1,schelhaas2,Rose,Ludwig}.

	As recently discussed in \cite{kappa} the constant that reflects the
	strength of
	mesonic effects, the so-called ``enhancement constant'' $\kappa $, enters
	into the description of nuclear Compton scattering at two stages. First,
	it describes the over-fulfillment of the dipole sum rule due to
	momentum-dependent potentials. Second, it is used to parameterize the
	additional term in the non-resonant amplitude due to Compton
	scattering by correlated proton-neutron pairs \cite{christillin}.
	Thus, in addition to defining the low-energy behaviour of the Compton
	amplitude, $\kappa $ is also present in the expression for the
	integrated photoabsorption cross section.
	The comparison of experimental data with the predictions on the grounds
	of dipole and quadrupole sum rules
	shows  that $\kappa $ here contributes approximately 30 per cent
	\cite{kappa}.

	In the present article we restrict ourselves to the consideration of
	heavy nuclei. We assume that for large $N$ and $Z$ the quasiclassical
	approximation can be applied, with $N$ and $Z$ being the numbers of
    neutrons and protons, respectively.
    We start from the Fermi gas model calculating the corrections to
    the corresponding nuclear wave function using perturbation theory
    with respect to $\pi $- and $\rho $-meson exchange.
    These corrections are important, because they determine an essential
    part of the nuclear correlation functions, giving a significant
    contribution to the mesonic part of the Compton amplitude.
    In the course of this calculation it was realized that three-body
    corrections are as important as two-body corrections.

    Using the wave function obtained by this method, we constructed the
    mesonic contribution to the Compton amplitude and studied its
    dependence on the photon energy and the scattering angle.
    We also considered the terms arising from mesonic currents, when a
    $\Delta $-isobar excitation is allowed as an intermediate state.

    The applicability of our results is restricted
    to photon energies less than approximately 150 MeV.
    This is due to our use of the
    meson-nucleon interaction in the static limit (nonrelativistic
    approximation).

\section{Compton amplitude and Correlator}
In \cite{martin1} it was suggested to write the total nuclear
Compton amplitude $T_A$ as a sum of three
contributions provided by different physical mechanisms
(see also \cite{martin2} and \cite{martin3}):
the collective nuclear excitations (Giant Resonances),
$T_{GR}(\omega ,\theta )$, the scattering by quasi-deuteron clusters,
$T_{QD}(\omega ,\theta )$, and the nucleon Compton scattering,
$T_{N}(\omega ,\theta )$,
where $\omega $ is the photon energy and  $\theta $ is
the scattering angle.
Here we will only deal with the first two of these.
When the photon energy $\omega $ is small enough (less than pion mass), it
is possible to extract from these two contributions their resonance
parts, which have non-zero imaginary parts at $\theta =0$.
Due to the optical theorem these imaginary parts are connected with the
photoabsorption cross section. With the use of a dispersion relation one
can obtain the real part to these resonance contributions in forward
direction. Phenomenological arguments have been used to extend this
resonance contributions to non-forward directions (see e.g. \cite{ziegler}).
Also a phenomenological description has been suggested
for the non-resonant contributions (seagull parts) (see e.g.
\cite{martin1}). Unfortunately recent experimental results show
discrepancies between this approach and the experimental data
\cite{helium-lund,
carbon-lund}.
The strong restrictions on the resonance parts due to dispersion theory
make them less model dependent than the non-resonant parts. Thus, in
order to achieve an agreement with the experiments, we investigate the
non-resonant parts with the use of microscopic considerations.

Due to the low-energy theorem, the Compton scattering amplitude at
$\omega =0$ is equal to its Thomson limit
\begin{equation}
	T_A(0,\theta )=-\,\vec \varepsilon _1\cdot \vec \varepsilon
	_2\;{{Z^2e^2}
	\over {AM}}    .
	\label{thomson}
\end{equation}
Here $\vec \varepsilon _1 $ and  $\vec \varepsilon _2 $
are the polarisation vectors of the incoming
and outgoing photons, respectively, $M$ is the nucleon mass and $e$ is the
proton charge, $e^2 = 1/137$.
The low-energy limit of the resonance part of the Compton amplitude can
be represented as
\begin{equation}
	R(0,\theta )=\,\vec \varepsilon _1\cdot \vec \varepsilon _2\;{{Ze^2}
	\over M}\,\,{N \over A}\,(1+\kappa )    ,
	\label{resonance}
\end{equation}
where $\kappa $ is the so-called enhancement constant. This form
(\ref{resonance}) is related to the dipole sum rule for the photoabsorption
cross section.

The non-resonant part can be represented as a sum of two
contributions: the kinetic seagull amplitude $B(\omega ,\theta )$
and the mesonic seagull
amplitude ${\cal M }(\omega ,\theta )$.
The kinetic seagull amplitude corresponds to the
Thomson scattering by individual nucleons inside the nucleus:
\begin{equation}
	B(\omega ,\theta )=-\,\vec \varepsilon _1\cdot
	\vec \varepsilon _2\;{{Ze^2} \over M}\,F_p(\vec \Delta )  ,
	\label{s2-1}
\end{equation}
where
\begin{equation}
	F_p(\vec \Delta )=\int {d\vec r\,}\rho _p(r)\,e^{i\,\vec \Delta \cdot \vec r}
	\label{s2-2}
\end{equation}
is the form factor belonging to the distribution $\rho
_p(r)$ of protons inside the nucleus
and $\vec \Delta $ is the momentum transfer, $\vec \Delta = \vec k_2
-\vec k_1 $,
with $\vec k_1$ and $\vec k_2$ being the momenta of the incoming and
outgoing photons, respectively.
Therefore, in order to obtain eq.\,(\ref{thomson}), it is necessary for
the mesonic seagull amplitude to satisfy the relation
\begin{equation}
	{\cal M }(0,0)=-\vec \varepsilon _1\cdot \vec \varepsilon _2\;{{ZN} \over
	{AM}}\,e^2\kappa   .
	\label{s2-6}
\end{equation}
Due to eq.\,(\ref{s2-6}) our calculation of $\kappa $, as
described in Section 3, can be compared with the value obtained from
experimental photoabsorption data. This provides a test of the
low-energy properties of the amplitude in our approach.

We consider mesonic contributions to Compton scattering in heavy nuclei
and apply a Fermi gas model of nuclear matter for the nuclear wave functions.
Such an
approach was also used in Ref. \cite{ericson1,ericson2} for the case of
forward direction.
There it was shown that in such an approach all nuclear degrees of
freedom can be taken into account by using a correlation function.
Such a correlator consists of a central and a tensor part. In a pure Fermi
model the tensor correlator is zero. However, the importance of tensor
correlations was already pointed out in \cite{ericson2}. In its effect on
the enhancement constant $\kappa $, for example, the central correlator
makes a comparatively small contribution due to the strong compensation
between the different diagrams. In \cite{ericson2} the results of
numerical nuclear matter calculations for the correlators were used and
compared to a similar calculation based on a variational principle
\cite{fantoni}.
In order to obtain analytical results for the
Compton amplitude and take into account not only the central part, but
also the tensor part of the correlator, we calculated the
corrections to the nuclear wave function,
using the interaction between nucleons due to $\pi $- and $\rho $-meson
exchange
as a perturbation.
It is clear that the accuracy of such an approach is limited. But we hope
to reproduce qualitatively most of the features of the amplitude.
Our analytical representation of the correlators allows us to study the
dependence of the amplitude on different parameters (nuclear radius, $N$
and $Z$), as well as its energy and angular distributions.

We start our investigations from the static Hamiltonian for the
pion-nucleon interaction (see e.g. \cite{ew})
\begin{equation}
	H_{\pi NN}=-{f \over {m}}\left( {\vec \sigma \cdot \vec \nabla }
	\right)\left( {\underline \tau \cdot \underline \phi } \right)   ,
	\label{s2-7}
\end{equation}
together with a minimal coupling to the photon field. In eq.\,(\ref{s2-7})
$m$ is the pion mass.
For all coupling constants we use the same notation and values as given in
\cite{ew}, in particular $f^2/(4\pi )$=0.08.
 In eq.\,(\ref{s2-7}) underlined symbols denote vectors in (cartesian) isospin
space, while an arrow indicates a vector in coordinate space.

The mesonic seagull amplitude ${\cal M }$ can be written in the following form:
\begin{equation}
	{\cal M }=\int {{{d\vec Q} \over {(2\pi )^3}}}\;S^{ij}\;T_{ij} .
	\label{s2-8}
\end{equation}
The contribution $T^{ij}_{(\pi )}$
 of $\pi $-meson exchange to $T_{ij}$, corresponding to the diagrams
shown in Fig.\,(\ref{fig1}), is given by:
\begin{eqnarray}
	T_{(\pi )}^{ij} &=& {{2e^2f^2} \over {m^2}}\;\left\{ {\matrix{{}\cr
{}\cr
{}\cr
}} \right.{{\varepsilon _1^i\varepsilon _2^j} \over {D_1}}
-2\;{{\varepsilon _1^iQ_2^j\;\vec \varepsilon _2\cdot \vec Q_2}
\over {D_1d_2}}-2\;{{\varepsilon _2^iQ_2^j\;\vec \varepsilon _1\cdot \vec Q_2}
	\over {D_2d_2}}   \nonumber  \\
	&+& 4\;{{Q_2^jQ_1^i\;\vec \varepsilon _1\cdot \vec Q_1\;\vec \varepsilon
_2\cdot
	\vec Q_2} \over {D_1d_1d_2}}
	-\;{{Q_2^jQ_1^i\;\vec \varepsilon _1\cdot \vec \varepsilon _2}
	\over {d_1d_2}}
	+\left( {\matrix{{i\leftrightarrow j}\cr
{\vec Q\leftrightarrow -\vec Q}\cr
}} \right)
	\left. {\matrix{{}\cr
{}\cr
{}\cr
}} \right\}   ,
	\label{b1}
\end{eqnarray}
where the following abbreviations have been used:
\begin{displaymath}
	D_{1,2}=\left( {\vec Q\pm \vec K} \right)^2+m^2-\omega ^2  \;,\quad
		\vec Q_{1,2}=\vec Q\pm {{\vec \Delta } \over 2}\;,\quad
	d_i=\vec Q_i^2+m^2    ,
\end{displaymath}
with $\vec K = (\vec k_1 + \vec k_2)/2$.
The correlator $S^{ij}$ entering into eq.\,(\ref{s2-8}) has the following
general form:
\begin{equation}
	S^{ij}=\sum\limits_{a\ne b} {\left\langle 0
	\right|\tau _a^{(-)}\tau _b^{(+)}\sigma _a^i\sigma _b^j\;}e^{i\;\vec Q\cdot
	(\vec x_b-\vec x_a)}\;e^{i\;\vec \Delta \cdot (\vec x_a+\vec x_b)/2}\left| 0
	\right\rangle   .
	\label{s2-9}
\end{equation}
Here the summation with respect to $a$ and $b$ is performed over all
nucleons, $\tau ^{(\pm )} = (\tau _1 \pm i \tau _2)/2$ are the isospin
raising and lowering operators.
The internal pion momenta were defined in such a way that photon and
pion momenta add up to $\vec Q + \vec \Delta /2$ in the left nucleon
vertex and $\vec Q - \vec \Delta /2$ in the right one.
This gave us the possibility to write the amplitude ${\cal M }$ in the form
(\ref{s2-8}), i.e. to extract the correlator (\ref{s2-9}).

The correlator $S^{ij}$ reveals some properties, which are interesting
for investigation. Here we will study its dependence on the nuclear
radius $R$, on $N$ and $Z$ and on the momentum transfer $\vec \Delta $. We
will do this for the simplest case, i.e within the framework of a pure
Fermi gas model.
Then the correlator only has a central part: $S^{ij} = S_C \delta ^{ij}$,
where
\begin{equation}
	S_C= -2 \int {{{d\vec p_1d\vec p_2}
	\over {(2\pi )^6}}}\int {d\vec x_1d\vec x_2}\,e^{-i(\vec x_1+\vec x_2)\vec
	\Delta /2}\,e^{i(\vec x_1-\vec x_2)((\vec p_1-\vec Q-\vec p_2)}   .
	\label{a1-1}
\end{equation}
The ranges of integration for the nucleon
momenta $\vec p _1$ and $\vec p _2$ are the proton and neutron
Fermi spheres, with the radii
\begin{displaymath}
	p_F^{(p)}=2.27\,m\,\left( {{Z \over A}}
	\right)^{1/3}\;,\quad p_F^{(n)}=2.27\,m\,\left( {{N \over A}}
	\right)^{1/3}\;,\quad A=N+Z .
\end{displaymath}
In eq.\,(\ref{a1-1}) the integrations with respect to $\vec x_i$ are taken
over a sphere with the radius $R = 1.2\,A^{1/3}\,$fm.
Straightforward integration gives
\begin{eqnarray}
   S_C&=& -{{R^6\left( {{{a+b} \over 2}} \right)}
   \over {6\pi ^2}}^6\sum\limits_{l=0}^\infty
   {(2l+1)P_l\left( X \right)}\times
	\nonumber \\
	& &\left\{ {\matrix{{}\cr
{}\cr
}} \right.\int\limits_c^2 {d\nu \,\nu (\nu -2)^2\left[ {\nu ^2+4\nu -3c^2}
\right]}\;g_1^{(l)}(\nu )g_2^{(l)}(\nu )
\nonumber   \\
& &+2(2-c)^3\int\limits_0^c
{d\nu \,\nu ^2}\,g_1^{(l)}(\nu )g_2^{(l)}(\nu )\left. {\matrix{{}\cr
{}\cr
}} \right\}
	\label{exact}
\end{eqnarray}
with $X=(\vec Q_1\cdot \vec Q_2)/(\left| {\vec Q_1} \right|\left| {\vec Q_1}
\right|)$, $a=p_F^{(p)}$, $b=p_F^{(n)}$ and $c=2(a-b)/(a+b)$.
The functions $g_i^{(l)}$ can be expressed via spherical Bessel
functions $j_l$,
\begin{equation}
	g_i^{(l)}(\nu )={{j_l(\nu R)Q_iR\,j_{l-1}(Q_iR)-j_{l-1}(\nu R)\nu
R\,j_l(Q_iR)}
	\over {(\nu R)^2-(Q_iR)^2}}     .
	\label{a1-4}
\end{equation}
Let us investigate the dependence of the correlator on the nuclear radius
$R$. For simplicity we consider the case $\vec \Delta =0$ and $N$=$Z$. Then we
have
\begin{eqnarray}
    \left. {S_C} \right|_{\scriptstyle {\vec \Delta =0,}\hfill\atop
  \scriptstyle {N=Z}\hfill}&\equiv &S_C^{(f)} = - {A \over 2}{{(Ra)^3}
	\over {3\pi }}\;\int\limits_0^2 {d\nu \;\nu ^2\left( {1-{\nu
	\over 2}} \right)^2\left( {1+{\nu  \over 4}} \right)\times }
	\nonumber \\
	& & \quad \quad \quad \int\limits_{-1}^{+1} {dX\;F^2(Ra\sqrt {q^2+\nu
	^2-2q\nu X})}  ,
	\label{a3}
\end{eqnarray}
with $q=Q/p_F$ and $p_F = p_F^{(p)} =  p_F^{(n)} = 1.8\,m $.
The function $F$ is given by
\begin{equation}
	F(x)={3 \over {x^2}}\left( {{{\sin x} \over x}-\cos x} \right)   .
	\label{form}
\end{equation}
The superscript $(f)$ for the correlator denotes that this result
corresponds to a finite nuclear radius.
Note that in a Fermi gas $Ra= (9\pi A/8)^{1/3}$.
If $R$ tends to infinity, one
obtains the asymptotic form of (\ref{a3}):
\begin{equation}
	S_C^{(0)}=-{A \over 2}\left( {1-{q \over 2}}
	\right)^2\left( {1+{q \over 4}} \right)\;\theta (2-q)     ,
	\label{a1}
\end{equation}
which is a well-known result (see e.g. \cite{Molinari}) of the Fermi model.
Here $\theta (x)$ is the step function.
In Fig.\,(\ref{fig3}), eq.\,(\ref{a3}) is compared with eq.\,(\ref{a1}) for
$A$=50 and $A$=200.
One can see that even in the case of $A$=50 the use of the asymptotic
form for the correlator is justified, even though in the region of very
small $Q$ the curves differ significantly.

We now consider the correlator $S_C$ from eq.\,(\ref{exact}), when
$\Delta $=0, $R$$\rightarrow $$\infty $ , but $N$$\ne $$Z$.
Then we have, for $N>Z$:
\begin{eqnarray}
	& &\tilde S_C^{(0)}=-{A \over 2}\;{{(a+b)^3} \over {(a^3+b^3)}}\,\left\{
{\matrix{{}\cr
{}\cr
}} \right.{{2a^3} \over {(a+b)^3}}\,\theta (b-a-Q)
	\label{asymnz} \\
	& &+{1 \over 8}\left( {1-{Q \over {a+b}}} \right)^2\left[ {2+{{Q^2-3(a-b)^2}
	\over {Q(a+b)}}} \right]\,\,\theta (a+b-Q)\,\theta (Q+a-b)\left.
{\matrix{{}\cr
{}\cr
}} \right\}  .
	\nonumber
\end{eqnarray}
Let us introduce the parameter $x=(N-Z)/A$. In Fig.\,(\ref{fig2}) the
quantity $\tilde S_C^{(0)}$
is shown as a function of $Q/p_F$ for different values of $x$.
One can see that even for large $x\sim 0.5$ the difference between $S_C^{(0)}$
and $\tilde
S_C^{(0)}$ is essential only in the region of very small $Q$. In our
further analytical calculations we will neglect this dependence and put
$N$=$Z$=$A/2$. In addition to this we will consider the limit of large
$R$, but $R\Delta \sim 1$.
Then it is easy to show that the correlator $S_C$ can be represented in
the following form:
\begin{equation}
	S_C=S_C^{(0)}\;F(R\Delta ) ,
	\label{deltadep}
\end{equation}
where $F$ and $S_C^{(0)}$ were defined in eqs.\,(\ref{form}) and
(\ref{a1}), respectively. For the corrections to the correlator, which
will be investigated in the next section, the dependence on the momentum
transfer $\vec \Delta $ is also determined by the form factor $F(R\Delta )$.

\section{Corrections to the Correlator}
As was mentioned in Section 2, in order to obtain the correct behavior of
the Compton amplitude, it is necessary to take into account not only the
central part of the correlator, but also its tensor part.
Therefore, in this section we calculate the correction to the wave
function. This can be done with the use of usual perturbation theory up
to second order. In this case we have to consider two-body and three-body
effects with respect to correlated nucleons.
Again all dependence of the correlator on the scattering angle is given
by the form factor $F(R\Delta )$, as pointed out in the last section.
So, let us consider eq.\,(\ref{s2-9}) in the case of $\vec \Delta $=0, which
without change of notation we will represent in the following form:
\begin{equation}
	S^{ij}=S_C\;\delta ^{ij}+S_T\;t^{ij}   ,
	\label{s2-10}
\end{equation}
where
\begin{equation}
	t^{ij}={{3Q^iQ^j} \over {Q^2}}-\delta ^{ij}  .
	\label{s2-11}
\end{equation}
The diagrams, which correspond to the two-body part of the correlator
correction, are shown in Fig.\,(\ref{fig4}).
Explicit evaluation of these diagrams leads to the following correction:
\begin{eqnarray}
	S_{(1)}^{ij}&=&-{{4Mf^2} \over {m^2}}V\int {{{d\vec p_1d\vec p_2}
	\over {(2\pi )^6}}}\;\left\{ {{{4Q^iQ^j}
	\over {\vec Q^2+m^2}}+{{2r^ir^j-\vec r\,^2\delta ^{ij}}
	\over {\vec r\,^2+m^2}}} \right\} \times
	\nonumber \\
	& &{{n(\vec p_1)n(\vec p_2)\left[ {1-n(\vec p_2+\vec Q)}
	\right]\left[ {1-n(\vec p_1-\vec Q)} \right]\,} \over {\vec Q\cdot \vec r}} ,
	\label{a4}
\end{eqnarray}
where
\begin{displaymath}
	\vec r=\vec p_1-\vec p_2-\vec Q
\end{displaymath}
and $V$ is the nuclear volume.
The function $n(\vec p) = \theta (p_F - |\vec p\,|)$ is the occupation number
in
the Fermi gas model.
Due to the symmetry of the integrand the product of four occupation
numbers can be omitted and one can replace
$[1-n(\vec p_2+\vec Q)] [1-n(\vec p_1-\vec Q)]$ by $[1-2n(\vec p_1-\vec
Q)]$.
Eq.\,(\ref{a4}) contributes to both, the central and tensor part of the
correlator. We can separate these two contributions with the help of the
identity
\begin{displaymath}
	4Q^iQ^j={4 \over 3}\vec Q^2(t^{ij}+\delta ^{ij})
\end{displaymath}
and the fact that due to the (spherically symmetric) integrations with
respect to the nucleon
momenta $\vec p_1$ and $\vec p_2$ the expression $(2r^ir^j-\vec r\,^2\delta
^{ij})$ can be replaced by
\begin{displaymath}
	{1 \over 3}\left\{ {{{3(\vec r\cdot \vec Q)-\vec r\,^2\vec Q^2}
	\over {\vec Q^2}}\;t^{ij}-\vec r\,^2\delta ^{ij}} \right\}  .
\end{displaymath}
Analytical integration of eq.\,(\ref{a4}) with respect to $\vec p_1$ and
$\vec p_2$ leads to the explicit formulae presented in the appendix.
It is important to realize that, in order to fulfill the requirement of
gauge invariance, one has to evaluate also the set of
three-body diagrams displayed in Fig.\,(\ref{fig5}).
As a result all contributions of the second order in the meson-nucleon
coupling are taken into account.
In accordance with the Feynman rules one obtains the following general
expression for the three-body contribution to the correlator:
\begin{eqnarray}
	S_{(2)}^{ij}&=&{{16Mf^2} \over {m^2}}V\int {{{d\vec p_1d\vec p_2}
	\over {(2\pi )^6}}}\;\left\{ {{{2Q^iQ^j}
	\over {\vec Q^2+m^2}}+{{2r^ir^j-\vec r\,^2\delta ^{ij}}
	\over {\vec r\,^2+m^2}}} \right\}
\times
	\nonumber \\
	& &{{n(\vec p_1)n(\vec p_2)n(\vec p_1+\vec Q)\,}
	\over {\vec Q\cdot \vec r}} .
	\label{threebody}
\end{eqnarray}
Here the product of four occupation numbers has also been omitted, as in
eq.\,(\ref{a4}).
The result of the explicit analytical integration for eq.\,(\ref{threebody})
can be found in the appendix.
As before, the correction (\ref{threebody}) contributes to the central
part of the correlator, as well as to its tensor part.
Thus, at this stage of our investigation the correlation function
$S^{ij}$ has the following form:
\begin{equation}
	S^{ij}=S_C^{(0)}\delta ^{ij}+S_{(1)}^{ij}+S_{(2)}^{ij}  .
	\label{fullcorr}
\end{equation}
Note that central and tensor parts of the corrections can be extracted
by using the definitions (\ref{s2-10}) and (\ref{s2-11}).
In Fig.\,(\ref{fig6}) the full central part of the correlator
as a function of $Q/p_F$ is shown, together with the
zeroth order approximation $S_C^{(0)}$ and the one, where only two-body
correlations have been taken into account.
It can be seen that almost over the whole range of $Q$ the
asymptotic form of the central correlator is modified,
although the overall shape remains the same.

Let us now pass on to the discussion of the tensor correlation and the
different contributions to it.
Fig.\,(\ref{fig7}) shows the tensor correlator with and without three-body
effects. Here a strong modification due to three-body correlations occurs at
comparatively low $Q$, where we found a strong damping that is in fact
essential for obtaining reasonable physical results.
A good test of the model developed so far is the calculation of the
enhancement constant $\kappa $.
With the use of eq.\,(\ref{s2-6}), together with eqs.\,(\ref{s2-8}) and
(\ref{b1}), we
obtain
\begin{equation}
	\kappa =-{1 \over Z}\,\left( {{{f^2} \over {4\pi }}} \right)\,\,{{16} \over
\pi }\,{M
	\over m}\,\int\limits_0^\infty  \,\,\left[ {{{S_C-2s^2S_T}
	\over {\left( {s^2+1} \right)^2}}-{{4s^2(S_C+2S_T)}
	\over {3\left( {s^2+1} \right)^3}}} \right]\,s^2ds  ,
	\label{kappa}
\end{equation}
where the functions $S_C$ and $S_T$ depend on $Q=ms$ and contain the two-
and three-body corrections, which are given in the appendix.
Our calculation is restricted by the use of the nonrelativistic
approximation. We, therefore, cannot apply our results at large $Q/p_F$.
One can see from Fig.\,(\ref{fig7}) that the contribution of the high-$Q$
region is not negligible. It is known that taking into account only the
one-pion exchange leads to a wrong behavior in the tensor part of the
nucleon-nucleon
potential at small distances (i.e. at large $Q$) (see e.g. \cite{ew}).
Thus, it is necessary to include $\rho $-meson exchange in the
calculation of the tensor part of the correlator.
This can be done \cite{brownrho} by replacing $S_T$ by:
\begin{equation}
   \tilde S_T=S_T\,\left[ {1-2\,{{q^2+m^2 / p_F^2}
   \over {q^2+m_\rho ^2 / p_F^2}}} \right]  ,
   	\label{s3-5}
\end{equation}
where $m_\rho $ is the $\rho $-meson mass
and the following relation for the $\rho NN$-coupling constant has been
used:
\begin{displaymath}
    \left( {{{f_\rho } \over {m_\rho }}} \right)^2\left( {{f \over m}}
    \right)^{-2}=2  .
\end{displaymath}
Taking into account the $\rho $-meson improves essentially the behaviour
of our correlator at $Q\sim p_F$ and higher (cf. Fig.\,(\ref{fig9})).
Unfortunately the
uncertainty connected with the choice of the upper limit in the integral
representation for $\kappa $, eq.\,(\ref{kappa}), still exists.
The integrand for the $s$-integration in (\ref{kappa})  is
shown in Fig.\,(\ref{fig15}).
We
estimate the accuracy of the result obtained for $\kappa $ to be of the
order of 20 per cent. Introducing a cut-off at $Q/p_F = 2.7$, we obtain
the numerical values for the different contributions to $\kappa $.
Using $S_C^{(0)}\delta ^{ij}$ as a correlator gives $\kappa _0$=0.2, in
agreement with \cite{ericson1}. With the full central part of the
correlator, one finds $\kappa _C$=0.15. The contribution of the
tensor part is $\kappa _T$=0.86. So, the total value for the
enhancement constant is $\kappa =\kappa _C + \kappa _T =1.01$.

Mind that all results for $\kappa $ obtained in this section have been
calculated for infinite nuclear matter with $N$=$Z$. In Fig.\,(\ref{fig2})
the influence of the difference between $N$ and $Z$ has been shown for
the lowest-order form of the correlator. It is also interesting to study
the enhancement constant $\kappa $, which does not depend on the momentum
$\vec Q$, as a function of $N$ and $Z$.
We will do that for the simplest case of only the $S^{(0)}_C$-contribution
to $\kappa $ (i.e. $\kappa _0$).
Let us put in eq.\,(\ref{kappa}) $S_T=0$, $S_C=\tilde S_C^{(0)}$ (from
eq.\,(\ref{asymnz})) and multiply the rhs of (\ref{kappa}) by $A/(2N)$ in
accordance with the definition (\ref{s2-6}).
After straightforward integration one obtains
\begin{equation}
	\kappa _0={{AM} \over {ZN}}\,{{f^2V}
	\over {6\pi ^4}}\left\{ {-ab+{1
	\over 4}\left( {a^2+m^2+b^2} \right)\,\ln \left[ {{{\left( {{{a+b}
	\over m}} \right)^2+1} \over {\left( {{{a-b} \over m}} \right)^2+1}}}
	\right]} \right\}  ,
	\label{a3-4}
\end{equation}
where $a=p_F^{(p)}$, $b=p_F^{(n)}$, as before.
In Fig.\,(\ref{fig22}) the $(N,Z)$-dependence of $\kappa _0$ is shown.

\section{Energy and Angular Dependence}
In the previous section we used the full central and tensor parts of the
nuclear correlator to calculate the
enhancement constant $\kappa $.
Let us pass now to the consideration of the energy dependence of the
amplitude (\ref{s2-8}). We start from the low-energy limit of the
amplitude and take into account terms of the order of $\omega ^2/m^2$.
Making the expansion of $T_{(\pi )}^{ij}$, eq.\,(\ref{b1}), with respect
to $\vec K$ and $\vec \Delta $ up to second order, contracting with
$\delta ^{ij}$ and $t^{ij}$ and, finally, averaging with respect to the
angles of $\vec Q$, we obtain within our accuracy
\begin{equation}
    {\cal M }={{Ae^2} \over {4M}}\,F(R\Delta )\left[ {-\kappa +\,{{\alpha
\,\omega ^2}
    \over {m^2}}\;\vec \varepsilon _1\cdot \vec \varepsilon _2+{\beta
    \over {m^2}}\;\left( {\vec \varepsilon _1\times \vec k_1} \right)\cdot
\left( {\vec
    \varepsilon _2\times \vec k_2} \right)} \right]  ,
	\label{alphabeta}
\end{equation}
where
\begin{eqnarray}
	\alpha &=&{{4Mf^2} \over {\pi ^2m}}\int {s^2ds}\,\left[ {\matrix{{}\cr
{}\cr
}} \right.(S_C+2S_T)\;{{10s^6+39s^4+60s^2+15} \over {15\,(1+s^2)^5}}
	\nonumber \\
	& &-2S_T\;{{3s^2+5} \over {5\,(1+s^2)^3}}\left. {\matrix{{}\cr
{}\cr
}} \right]
	\label{alpha}
\end{eqnarray}
and
\begin{equation}
	\beta ={{4Mf^2} \over {\pi ^2m}}\int {s^2ds}\,\left[
{(S_C+2S_T)\;{{2s^4-5s^2-3}
	\over {3\,(1+s^2)^4}}+2S_T\;{{1-s^2} \over {\,(1+s^2)^3}}} \right]  .
	\label{beta}
\end{equation}
The constants $\alpha $ and $\beta $ can be interpreted
\cite{ericson1,martin1} as the modification due to medium effects of the
electromagnetic polarizabilities for the free nucleon, measured in units
of $\zeta $=$e^2/(4Mm^2)$=$7.5\times 10^{-4}$fm$^3$. Including this coefficient
we
obtain the following numerical results:
$\bar \alpha $=$\zeta \alpha $=$-$3.23$\times 10^{-4}$fm$^3$ and
$\bar \beta $=$\zeta \beta $=1.73$\times 10^{-4}$fm$^3$.
The use of only $S_C^{(0)}\delta ^{ij}$ leads to
$\bar \alpha _0$=$-$4.5$\times 10^{-4}$fm$^3$ and
$\bar \beta _0$=1.95$\times 10^{-4}$fm$^3$.
The sum $\bar \alpha $+$\bar \beta $=$-$1.5$\times 10^{-4}$fm$^3$,
which is obtained from the consideration of forward direction scattering,
gives a correction of the order of 10 per cent to the corresponding value for
the free
nucleon. The polarizabilities contribute to the backward scattering
amplitude in the combination $\bar \beta $$-$$\bar \alpha
$=5$\times 10^{-4}$fm$^3$. Here the relative correction is essentially larger
than in the case of forward direction.

Note that the expression in brackets in eq.\,(\ref{alphabeta}) does not
contain an electric quadrupole contribution. Some E2 strength occurs
due to the expansion of the form factor $F(R\Delta )$. However, this
contribution explicitely cancels with its resonance counterpart, in
accordance with the low-energy theorem.

It is interesting to compare the exact energy dependence of our amplitude
with its low-energy form (\ref{alphabeta}).
In Fig.\,(\ref{fig13}) this dependence is shown for the cases of forward
and backward scattering. One can see that up to an energy of 100 MeV all
energy dependence is reproduced by eq.\,(\ref{alphabeta}). We can,
therefore, apply the low-energy form of the amplitude at all scattering
angles.

\section{Virtual $\Delta $-excitation}
Let us consider now the contribution of a virtual $\Delta $-isobar
excitation to the mesonic part of the Compton amplitude.
General properties of such an effect were already discussed in
\cite{Arenh}. Here we have the possibility to consider this contribution
in more detail with the use of the correlation function obtained in
Section 3.
In the static limit the corresponding Hamiltonians, which determine the
interaction, are of the form  (see e.g. \cite{ew})
\begin{equation}
	H_{\gamma N\Delta }=-{{ef_{\gamma N\Delta }}
	\over m}\,\vec S^+\cdot \left( {\vec \nabla \times \vec A} \right)\,\,T_z^+
	\label{s5-1}
\end{equation}
and
\begin{equation}
	H_{\pi N\Delta }=-{{f_\Delta } \over m}\,\vec S^+\cdot
	\vec \nabla \underline T^+\underline \phi  \,\, ,
	\label{s5-2}
\end{equation}
with the hermitian conjugate to be added in both cases. Here
$\vec S$ and $\underline T$ are the 1/2-to-3/2 transition operators in
spin space and isospin space, respectively.
Evaluating the diagrams given in Fig.\,(\ref{fig8}) on this basis, one
obtains the following expression for the isobar contribution to the
tensor $T^{ij}$ entering into eq.\,(\ref{s2-8}):
\begin{equation}
	T_{(\Delta )}^{ij}=\eta \,\left[ {{{h_1^il_2^j+l_1^ih_2^j}
	\over {D_1}}+{{Q_1^i} \over {d_1}}\,\left[ {\vec \varepsilon _1\times (\vec
k_2\times
	\vec \varepsilon _2)-\,\vec \varepsilon _2\times (\vec k_1\times
	\vec \varepsilon _1)} \right]^j} \right]   ,
	\label{deltatij}
\end{equation}
with
\begin{displaymath}
     \vec h_a=\vec \varepsilon _a-{{2(\vec Q_a\cdot \vec \varepsilon _a)\vec
Q_a}
     \over {d_a}}\;,\quad \vec l_a=(\vec Q+\vec K)\times (\vec k_a\times
     \vec \varepsilon _a)
\end{displaymath}
and the coefficient
\begin{displaymath}
     \eta =-{{8\Omega e^2f_\Delta f_{\gamma N\Delta }f}
     \over {9(\omega ^2-\Omega ^2)m^3}}   .
\end{displaymath}
In (\ref{deltatij}) the same abbreviations have been used as in eq.\,(\ref{b1})
and, in addition, $\Omega = M_{\Delta } - M$ is the mass
difference between the $\Delta $-isobar and the nucleon.
Also, the symmetry with respect to the substitution
$\vec Q \rightarrow -\vec Q$, which is due to the integration in eq.\,
(\ref{s2-8}), has been used to bring $T^{ij}_{(\Delta )}$ into the form
(\ref{deltatij}).
As before, we consider the low-energy limit of the isobar contribution to
the amplitude, which again can be represented in the form
(\ref{alphabeta}). In this procedure we find $\alpha _{\Delta }$=0 and
\begin{equation}
	\beta _\Delta ={{16M\,f_\Delta f_{\gamma N\Delta }f}
	\over {27\pi ^2\Omega }}\,\,\int {s^2ds\,}{{2s^2S_T+(s^2-3)S_C} \over
	{(s^2+1)^2}}  .
	\label{betadelta}
\end{equation}
Inserting the numerical values for the coupling constants $f_{\Delta
}$=2$f$ and $f_{\gamma N\Delta }$=0.35, we obtain $\beta _{\Delta
}$=0.09 and, correspondingly, for the contribution of the $\Delta
$-isobar to the nucleon magnetic polarizability in the usual units
$\bar \beta _{\Delta }$=0.7$\times 10^{-4}$fm$^3$.
As before, the energy dependence of the amplitude ${\cal M }_{\Delta }$
discussed in this
section is well described by its low-energy limit. At energies up to the
pion mass ${\cal M }_{\Delta }$ is less than 0.1 in units of $Ae^2/(4M)$.

\section{Conclusion}
In this paper we have tried to understand qualitatively the influence of
different correlations inside heavy nuclei on the mesonic contribution to
Compton scattering.
Our model approach gives us the possibility to perform a large part of
the calculation analytically and, as a result, to understand in more
detail the interplay between the different physical contributions as well
as  the influence of various parameters.
We started from a simple Fermi gas model and then
investigated the correlations, which appear in such a gas due to $\pi $- and
$\rho
$-meson exchange between the nucleons.
In the calculation of these corrections, three-body diagrams have been
found to give a contribution of the same order of magnitude as two-body
diagrams.
For example, strong compensations between two- and three-body
contributions to the correlator  occur in the region of relatively small
pion momenta.

We obtained that the tensor part of the correlation function has as
important an effect on the Compton amplitude as its central part. In
particular it increases essentially the enhancement constant $\kappa $.
Such nuclear correlations have also been found to be significant in the
modification of the nucleon electromagnetic polarizabilities via mesonic
currents.
In addition we found that it is not unnatural to introduce, in the
analysis of experimental data, a rather high modification of $\bar
\beta $$-$$\bar \alpha $.

The contribution of the virtual $\Delta $-isobar to the magnetic
polarizability has been found to be less important than that of the
purely pionic amplitude.

\section*{Acknowledgement}
We are grateful to A.I. L'vov and M. Schumacher for useful comments and
stimulating discussions. A.I.M. wishes to thank II.
Physikalisches Institut, University of G\"ottingen, for the warm
hospitality during the stay, when this work was finished.
M.T.H. is most grateful to the Budker Institute, Novosibirsk, for the
kind hospitality accorded him during his stay, when part of this work was
done.

\section*{Appendix: Analytical expressions for the central and tensor
parts of the correlator}
Here we present the explicit expressions for the correlator corrections,
which have been used to obtain the figures shown in Section 3.
The two-body correction $S^{ij}_{(1)}$, eq.\,(\ref{a4}), leads to the following
additional
terms in the correlator:
\begin{equation}
	S_C^{(1)}=\;{A \over 2}C\;\left[ {G_1\theta (q-2)+G_2\theta (2-q)} \right]
	\label{AA1}
\end{equation}
for the central part and
\begin{eqnarray}
	S_T^{(1)}&=&{A \over 2}\;C\;\left\{ {\matrix{{}\cr
{}\cr
{}\cr
}} \right.\theta (q-2)\left[ {G_1+\int\limits_{q-1}^{q+1} {w(q,x)\,dx}} \right]
	\nonumber \\
	&+&\theta (2-q)\left[ {G_2+\left( {\int\limits_1^{q+1}
	{+\int\limits_1^{\left| {q-1} \right|} {}}} \right)w(q,x)dx} \right]\left.
{\matrix{{}\cr
{}\cr
{}\cr
}} \right\}
	\label{AA2}
\end{eqnarray}
for the tensor part of the correlator.
In (\ref{AA1}) and (\ref{AA2}) we used the following abbreviations:

The functions $G_1$ and $G_2$ are of the form
\begin{eqnarray}
	G_1&=&{{11} \over {60}}+{{q^2} \over {120}}+\left[ {-{1 \over {15q}}+{q
	\over {12}}-{{q^2} \over {24}}+{{q^4} \over {480}}} \right]\ln \left| {2-q}
\right|
	\nonumber \\
	&+&\left[ {{{q^2} \over {12}}-{{q^4} \over {240}}}
	\right]\ln (q)+\left[ {{1 \over {15q}}-{q \over {12}}-{{q^2} \over
{24}}+{{q^4}
	\over {480}}} \right]\ln (q+2)
	\label{AA4}
\end{eqnarray}
and
\begin{eqnarray}
	G_2&=&{{29} \over {240}}q-{{q^3} \over {320}}+\left[ {{1 \over {15q}}-{1
	\over {16}}+{{q^2} \over {96}}-{{q^4} \over {1280}}} \right]\ln (2-q)
	\nonumber \\
	&-&\left[ {{2 \over {15q}}+{q \over 6}} \right]\ln (2)+\left[ {{1
	\over {15q}}+{1 \over {16}}-{{q^2} \over {96}}+{{q^4} \over {1280}}}
	\right]\ln (2+q)  .
	\label{AA5}
\end{eqnarray}
The integrand $w(q,x)$ is given by
\begin{eqnarray}
	w(q,x)&=&\;{1 \over {32q}}\left[ {1-\left( {{{q^2+x^2-1}
	\over {2xq}}} \right)^2} \right]Ê\times
	\nonumber \\
	& &\left( {x(x^2+1)-{1 \over 2}(x^2-1)^2\ln \left| {{{x+1}
	\over {x-1}}} \right|} \right)  .
	\label{AA3}
\end{eqnarray}
The external factor is $C=(3Mf^2p_F)/(\pi ^2m^2)$.

It is clear that the main difference between the calculation of two- and
three-body correlations occurs in the structure of the product of
occupation numbers $n(\vec p)$
(cf. eqs. (\ref{a4}) and (\ref{threebody}))
It is convenient to perform some of the integrations in (\ref{threebody})
by representing the integrand as the one for the two-body
contributions plus some additional terms. This procedure leads to the
following expression for the correction $S_{(2)}^{ij}$ due to three-body
graphs:
\begin{equation}
	S_{(2)}^{ij}=2\,S_{(1)}^{ij}+\,\delta S^{ij}-\,{{2CA}
	\over 3}\,{{q^2\left( {G_2-G_1} \right)}
	\over {q^2+\gamma ^2}}\,\left( {t^{ij}+\delta ^{ij}} \right)\,\;\theta
	(2-q)   ,
	\label{AA6}
\end{equation}
with
\begin{eqnarray}
	\delta S^{ij}&=&-\,{{CA} \over 3}\left\{ {\matrix{{}\cr
{}\cr
{}\cr
}} \right.\left( {\,{{4q^2} \over {q^2+\gamma ^2}}-1}
\right)\,G_1\,\,(\delta ^{ij}+t^{ij})
   \nonumber   \\
& &+\,\int\limits_0^2 {x\,dx}\left( {1-{x
\over 2}} \right)^2\left( {1+{x \over 4}} \right)\times
	\nonumber \\
	& &\left[ {\matrix{{}\cr
{}\cr
{}\cr
}} \right.\left\{ {\ln \left| {{{q+x} \over {q-x}}} \right|-\ln \left[
{{{(q+x)^2+\gamma ^2}
\over {(q-x)^2+\gamma ^2}}} \right]} \right\}\,{{\gamma ^2\left( {\delta
^{ij}+t^{ij}}
\right)} \over {6q\left( {x^2+\gamma ^2-q^2} \right)}}
	\nonumber \\
	& &+{1 \over {2q^2}}\,\left\{ {1-{{\gamma ^2+q^2x^2}
	\over {4qx}}\,\ln \left[ {{{(q+x)^2+\gamma ^2} \over {(q-x)^2+\gamma ^2}}}
	\right]} \right\}\;t^{ij}\left. {\matrix{{}\cr
{}\cr
{}\cr
}} \right]\left. {\matrix{{}\cr
{}\cr
{}\cr
}} \right\}  .
	\label{AA7}
\end{eqnarray}
Here $\gamma = m/p_F$.
The coefficients of $\delta ^{ij}$ and $t^{ij}$ in  eq.\,(\ref{AA6}) are
the functions $S_C^{(2)}$ and $S_T^{(2)}$ used in Section 3.

\newpage
\section*{Figure Captions}
   	\begin{abb}
   	\protect\label{fig1}
	    Typical diagrams contributing to $T^{ij}_{(\pi )}$. The wavy
	    lines denote photons and dashed lines denote pions. The
	    amputation indicates that $T^{ij}$ contains only the nucleon vertices, but
	    not its wave functions.
	\end{abb}

   	\begin{abb}
   	\protect\label{fig3}
	    Comparison of the exact correlator $S_{C}^{(f)}$ with its
	    asymptotic form $S_{C}^{(0)}$ obtained for $R \rightarrow \infty $
	    (curve a)). In $S_{C}^{(f)}$
	    $A$ is set to 50 (curve b)) and 200 (curve c)).
	\end{abb}

   	\begin{abb}
   	\protect\label{fig2}
	   Dependence of $\tilde S_C^{(0)}$ on $Q/p_F$ for different $x=(N-Z)/A$
	   in units of $NZ/A$. The values used for $x$ are a) $x=$0.5, b)
	   $x=$0.25, c) $x=$0.1 and d) $x=$0.
	\end{abb}

   	\begin{abb}
   	\protect\label{fig4}
	    Set of diagrams, from which the two-body correction to the correlator
	    is extracted. Here the same symbolic abbreviation is used as in Fig.\,
	    (\ref{fig1}). The nucleon spin projections are denoted by $\lambda _i$.
	\end{abb}

   	\begin{abb}
   	\protect\label{fig5}
	    Three-body diagrams yielding an additional correction
	    to the
	    correlator, which is of the same order in $f/m$ as the two-body
	    diagrams shown in Fig.\,(\ref{fig4}). The notations are the same as
	    in Fig.\,(\ref{fig1}) and  Fig.\,(\ref{fig4}).
	\end{abb}

   	\begin{abb}
   	\protect\label{fig6}
	    Central part of the correlator a) with three-body corrections, b) without
	    three-body corrections and c) Fermi correlator
	    without any corrections.
	\end{abb}

   	\begin{abb}
   	\protect\label{fig7}
	    Tensor part of the correlator a) with three-body corrections and b)
without
	    three-body corrections.
	\end{abb}

   	\begin{abb}
   	\protect\label{fig9}
	    Comparison of the tensor part of the correlator a) with and b) without the
$\rho
	    $-meson contribution.
	\end{abb}

   	\begin{abb}
   	\protect\label{fig15}
	    Integrand for the calculation of $\kappa $ shown as a function of
	    $Q/m$: a) the full integrand, b) contribution of the central part
	    and c) contribution of the tensor part of the correlator.
	\end{abb}

   	\begin{abb}
   	\protect\label{fig22}
	    Dependence of $\kappa _0$ on $N$ and $Z$.
	\end{abb}

    \begin{abb}
   	\protect\label{fig13}
	    Energy dependence of the amplitude ${\cal M }_\pi $. The exact dependence
	    (full line) is compared with the low-energy form (dashed line) for
	    scattering a) in forward direction and b) in backward direction.
	    ${\cal M }_\pi $ is shown in units of $-Ae^2/(4M)$.
	\end{abb}

   	\begin{abb}
   	\protect\label{fig8}
	    Set of diagrams, which arises from allowing a $\Delta $-isobar
	    excitation as an intermediate state in the diagrams displayed in
	    Fig.\,(\ref{fig1}).
	\end{abb}

\end{document}